\documentclass[12pt]{iopart}

\usepackage{iopams}
\usepackage{epsfig}
\usepackage{epstopdf}
\usepackage{graphicx}
\usepackage{lineno}

\begin{document}

\title{Periodic dynamics of fermionic superfluids in the bcs regime}

\author{A Roy$^1$, R Dasgupta$^2$\footnote{Current address: Asia Pacific Center for Theoretical Physics (APCTP), Pohang, Gyeongbuk 790-784, Korea}, S Modak$^3$, A Das$^3$, and K Sengupta$^3$}
\address{$^1$TCMP division, Saha Institute of Nuclear Physics,
1/AF Bidhannagar, Kolkata-700064, India \\
$^2$ S. N. Bose National Centre for Basic Sciences, 3/JD Bidhannagar,
Kolkata-700098, India.\\
$^3$ Theoretical Physics Department, Indian Association
for the Cultivation of Science, Kolkata-700032, India.}
\ead{daneel@utexas.edu}

\begin{abstract}
We study the zero temperature non-equilibrium dynamics of a fermionic superfluid in
the BCS limit and in the presence of a drive leading to a time
dependent chemical potential $\mu(t)$. We choose a periodic driving
protocol characterized by a frequency $\omega$ and compute the
fermion density, the wavefunction overlap, and the residual energy
of the system at the end of $N$ periods of the drive. We demonstrate
that the BCS self-consistency condition is crucial in shaping the
long-time behaviour of the fermions subjected to the drive and
provide an analytical understanding of the behaviour of the fermion
density $n_{{\mathbf k}_F}$ (where ${\mathbf k}_F$ is the Fermi momentum vector) after a drive
period and for large $\omega$. We also show that the momentum
distribution of the excitations generated due to such a drive bears
the signature of the pairing symmetry and can be used, for example,
to distinguish between s- and d-wave superfluids. We propose
experiments to test our theory.

\end{abstract}

\maketitle

\section{\sc Introduction:}
\label{sec:intro}

Ultracold atoms provide us with a useful test bed for studying
equilibrium and non-equilibrium properties of interacting many-body
systems. The initial focus in these systems has been largely on
bosonic atoms; in particular, the realization and the study of
properties of Bose-Einstein condensates (BECs) has been the prime
subject of investigation in the first few years of experimental
studies on such systems \cite{rev1}. In contrast, studies of
fermionic atoms have gained momentum much later \cite{jin1,rev2}.
The main experimental obstacle in studying many-body effects in
fermionic atoms has been the realization of sufficiently low
temperature so as to obtain a gas of quantum degenerate fermions
with $T \le T_F \sim \hbar^2 n_0^{2/3}/(k_B m)$, where $m$ is the
mass of the atoms and $n_0$ is their density, $T_{F}$ is the Fermi
temperature of the gas and $k_B$ is the Boltzmann constant. However,
recent experiments have made significant progress in this direction
and it has been possible to observe the crossover from classical to
quantum behaviour in fermionic gases \cite{exp1}. The formation of
Fermi superfluids, which is an interesting many-body phenomenon in
its own right \cite{rev3}, required lower temperature and stronger
interactions. It was soon realized that the latter can be achieved
by utilizing the Feshbach resonance phenomenon which allows for
tuning of both the strength and the sign of the interaction between
the fermions. A major hindrance in realizing such strong
interactions for bosonic atoms has been three-body losses; in
contrast, such losses are minimal for fermionic atoms due to the
Pauli exclusion principle. This allows for the possibility of stable
fermionic condensates with strong inter-particle interaction which
acts as a test bed for studying Fermi superfluids and, in
particular, the BCS-BEC crossover in these systems. Several recent
experiments have verified this phenomenon by numerous measurements
in both the BCS and the BEC side of the crossover \cite{exp2}.

The dynamical properties of Fermi superfluids have also received
theoretical and experimental attention in the recent past. On the
experimental side, there have been several studies such as probing
the expansion of Fermi superfluids after a sudden release of the
trap potential \cite{ohara1}, measurement of collective excitations
of these superfluids \cite{barnstein1}, measurement of the
superfluid gap by radio-frequency (RF) spectroscopy \cite{chin1},
and observation of vortex dynamics \cite{zwierlien1}. On the
theoretical side, several studies were made to study the equilibrium
and near-equilibrium properties of these systems. In particular,
early studies concentrated on understanding the crossover phenomenon
by approaching it from the BCS side \cite{leggett1}. These have been
later supplemented by inclusion of more sophisticated diagrammatic
techniques over the BCS mean-field theory \cite{pieri1}, study of
the effect of presence of a trap potential \cite{petrov1}, inclusion
of bosonic molecular degree of freedom in the BCS Hamiltonian
\cite{griffin1}, and use of quantum Monte Carlo methods
\cite{carlson1}. Later works focused on non-equilibrium aspects of
these systems based on hydrodynamic approach for studying low-lying
collective excitations \cite{menotti1}, vortex dynamics \cite{vor1},
quench dynamics across a BCS-BEC crossover \cite{rajdeep1}, and
properties of dynamic structure factors of these superfluids in the
weak-interaction regime \cite{minguzzi1}.

Non-equilibrium dynamics of closed quantum systems have recently
received a lot of theoretical attention due to the possibility of
realizing such dynamics in ultracold atom systems
\cite{Kris-RMP,Dziarmaga-rev,Damski-LZ,Zurek,anatoli,diptiman,
shreyoshi,rigol,arnab1,anatoli1,ks1,ks2,ar1}. Most of such
studies in this direction have concentrated on bosonic or spin
Hamiltonians realized by bosonic ultracold atoms in optical lattices
\cite{rigol,anatoli1,ks1,ks2}. In particular, experimental
realizations of Ising-like spin model \cite{bakr1} and Bose-Hubbard
model \cite{greiner1} have provided impetus to such theoretical
studies. More recently, concrete experiments were carried out on the
dynamics of bosons near the superfluid-insulator transition
\cite{bakr2}. The results of such experiments are in qualitative
agreement with theoretical studies on such systems \cite{ks2}.
Similar attempts of experimental realization of the Ising model have
recently been undertaken in trapped ion systems \cite{kim1,fr1}.
However, such studies have not been carried out extensively on
fermionic atoms in the superfluid state.

In this work, we study the response of a fermionic superfluid in the
BCS regime to a periodic drive. We choose a specific driving
protocol which leads to a time-dependent periodic chemical potential
for the fermions characterized by a frequency $\omega$: $\mu(t)=
\mu_0 + \mu_a \sin(\omega t)$. We note that such periodic drives are
known to lead to a host of interesting phenomena in quantum systems.
For example, it has been observed that coherent periodic driving in
a class of integrable quantum many-body systems can give rise to
novel quantum phenomena like dynamical many-body freezing, where
non-monotonic freezing behaviour (with respect to the driving
frequency) is observed\ \cite{arnab1}. A variant of this phenomenon
has also been predicted for ultracold bosons in optical lattices
\cite{pekker1}. The aim of the present work is to study the effect
of such a drive on superfluid fermions.

The key results that we obtain from such a study are the following.
First, we show that the BCS self-consistency condition plays a
crucial role in shaping the response of such superfluids to the
periodic drive and hence establish that the dynamics of fermionic
superfluids will be fundamentally different from those of integrable
systems such as Ising or Kitaev models which can be described by
Bogoliubov-like Hamiltonians without the self-consistency condition.
We demonstrate this by computing the fermion density (which can be
easily related to the magnetization of the Ising and Kitaev models)
which displays oscillatory behaviour as a function of time for the
Ising system and approaches a constant at long time for the
self-consistent BCS system. We also derive an analytical formula for
the $\omega$ dependence of the fermion density $n_{{\mathbf k}_F}$ (or
equivalently magnetization $m_{{\mathbf k}_F}$) at the gap edge (where ${\mathbf k}_F$
is the Fermi momentum vector) after a complete drive cycle and in the limit
of large drive frequency. Second, we compute the wavefunction
overlap (and hence the defect density) and the residual energy of
the systems after single and multiple cycles of the drive and
discuss the dependence of these quantities on $\omega$. Finally, we
compute the momentum distribution of the excitations created due to
the drive at the end of one drive cycle and show that such a
distribution depends on the pairing symmetry of the fermionic
superfluid. Thus we demonstrate that the dynamic response of these
superfluid may prove to be a useful tool for determining its pairing
symmetry.

The plan of the rest of paper is as follows. In\ 
\sref{formalism}, we introduce the model and the corresponding BCS
mean-field equations and provide explicit expressions for the
observables that we shall compute. In\ \sref{numerics}, we
present numerical results for several observables such as the defect
density, its momentum distribution, and the residual energy at the
end of a drive cycle and discuss their properties. This is followed
by an analytical treatment of the self-consistent problem in\ 
\sref{analytic} where we obtain an analytical expression for the
$\omega$ dependence of $m_{{\mathbf k}_F}$ at high $\omega$. We provide a
discussion of our work and suggest possible experiments to test our
theory in\ \sref{discuss} and provide some calculational details
in the appendix.

\section{Formalism}
\label{formalism}

In this section, we introduce the formalism and define the main
physical observables which we shall compute numerically. The
Hamiltonian for a gas of interacting ultracold fermions in a shallow
square optical lattice at $T=0$, in the absence of any drive, is given by
\begin{eqnarray}
H(t) &=& \sum_{{\mathbf k}\sigma} \left[ \epsilon_{\bf
k}-\mu_0\right]\hat{c}^\dagger_{{\mathbf k}\sigma}\hat{c}_{{\mathbf k}\sigma}
\nonumber\\
& &-g \sum_{{\mathbf k}, {\mathbf k'},{\mathbf k''}} \hat{c}^\dagger_{{\mathbf k}+{\bf
k''}\uparrow} \hat{c}^\dagger_{{\mathbf k'}-{\mathbf k''}\downarrow} \hat{c}_{{\bf
k'}\downarrow} \hat{c}_{{\mathbf k}\uparrow}. \label{fermiham}
\end{eqnarray}
Here $\hat{c}_{{\mathbf k}\sigma}$ represent the annihilation operators for
fermions of momentum ${\mathbf k}$ and spin $\sigma=\{\uparrow,
\downarrow\}$. The first term represents the kinetic energy of the
fermions, and the second term the four-fermion interaction energy
with amplitude $g>0$ which represents attractive interaction between
the fermions. Here $\epsilon_{\mathbf k}= -2J \sum_{i}\cos(k_i)$ is the
band energy spectrum for the fermions, the index $i$ takes values
$x$ and $y$ for $d=2$ or $x$, $y$, and $z$ for $d=3$, and $\mu_0$ is
the chemical potential. In the rest of this work, we shall assume
that the trap potential is slowly-varying so that a locally constant
chemical potential $\mu_0 = \epsilon_{F}$ (where $\epsilon_F$ is the Fermi energy \textit{viz.} the energy at \
the Fermi momentum vector ${\mathbf k}_F$) can be used to describe the fermions in the trap. In the BCS regime and at zero temperature,
the ground state of the fermions is a superfluid whose excitations
can be described by the BdG equations
\begin{eqnarray}
E({\mathbf k})\left(\begin{array}{c} u_{\mathbf k} \\ v_{\mathbf k}
\end{array} \right) &=& \left(
\begin{array}{cc}
(\epsilon_{\mathbf k}-\mu_0) & \Delta({\mathbf k})\\
\Delta^{\ast}({\mathbf k}) & -(\epsilon_{\mathbf k}-\mu_0)
\end{array} \right) \left(\begin{array}{c} u_{\mathbf k} \\ v_{\mathbf k}
\end{array} \right), \nonumber\\ \label{bdg1}
\end{eqnarray}
where $u_{\mathbf k}$ and $v_{\mathbf k}$ are the amplitudes of the particle
and the hole in a BdG quasiparticle and are related to the BCS
wavefuntion by
\begin{eqnarray}
|\psi\rangle &=& \prod_{\mathbf k} (u_{\mathbf k} + v_{\mathbf k} \hat{c}_{\bf
k}^{\dagger} \hat{c}_{{\bf -k}}^{\dagger}) |0\rangle. \label{bcswav}
\end{eqnarray}
The pair-potential $\Delta({\mathbf k})$ depends on the pairing symmetry
and is given by
\begin{eqnarray}
\Delta({\mathbf k}) &=& \Delta_0, \quad {\rm s-wave }, \nonumber\\
\Delta({\mathbf k}) &=& \Delta_0 [\cos(k_x) -\cos(k_y)],  \quad {\rm
d}_{x^2-y^2}{\rm -wave}. \label{pp}
\end{eqnarray}
In the rest of this work, we shall mostly consider s-wave pairing
except while discussing momentum distribution of the defect density
in\ \sref{numerics}, where we shall discuss other pairing
symmetries. Our analysis, which will be detailed in this section,
can be easily generalized to other pairing symmetries. For the rest
of this work, we set $\hbar=1$.

For s-wave pairing, the pair potential satisfies the
self-consistency relation
\begin{eqnarray}
\Delta_0 &=& g \sum_{\mathbf k} u_{\mathbf k}^{\ast} v_{\mathbf k}.
\label{self1}
\end{eqnarray}
\ \Eref{bdg1} and \eref{self1} admit the well-known BCS solution
\begin{eqnarray}
E({\mathbf k})&=& \pm \sqrt{(\epsilon_{{\mathbf k}}-\mu_0)^2+|\Delta_0|^2},
\nonumber\\
u^{\rm eq}_{\mathbf k}  &=& \frac{1}{\sqrt 2}
\left[ 1+ \frac{(\epsilon_{\mathbf k} -\mu_0)}{E({\mathbf k})}
\right]^{1/2},\nonumber\\
v^{\rm eq}_{\mathbf k}&=&\frac{1}{\sqrt 2}\left[ 1- \frac{(\epsilon_{\mathbf k} -\mu_0)}{E({\mathbf k})}
\right]^{1/2} e^{-i\phi_0}. \label{eqsol}
\end{eqnarray}
Here, $\phi_0$ is the phase of $\Delta_0$. We now introduce a 
time-dependent drive, $\mu(t)=\mu_0 +\mu_a
\sin(\omega t)$, so that $\mu_a, \,  \omega \ll J $. This can be
achieved in typical experimental systems by introducing an
additional time-dependent harmonic trap potential which is
sufficiently broad so as to allow for a uniform fermion density. In
this regime, the response of the system to the drive can be
described by the time-dependent Bogoliubov de-Gennes equation given
by
\begin{eqnarray}
i \partial_t \left(\begin{array}{c} u_{\mathbf k}(t) \\ v_{\mathbf k}(t)
\end{array} \right) &=& \left(
\begin{array}{cc}
(\epsilon_{\mathbf k}-\mu (t)) & \Delta({\bf k;t})\\
\Delta^{\ast}({\mathbf k};t) & -(\epsilon_{\mathbf k}-\mu (t))
\end{array} \right) \nonumber\\
& &\times \left(\begin{array}{c} u_{\mathbf k}(t)\\ v_{\mathbf k}(t)
\end{array} \right), \label{bdg2}
\end{eqnarray}
together with the self-consistency condition which, for s-wave
pairing, reads
\begin{eqnarray}
\Delta({\mathbf k};t) &\equiv& \Delta(t) = g \sum_{\mathbf k} u^{\ast}_{\bf
k}(t) v_{\mathbf k}(t).\label{tdsc1}
\end{eqnarray}
In the rest of this work, we consider the system to be in the
superfluid ground state at $t=t_i$ with $(u_{\mathbf k}(t_i), v_{\bf
k}(t_i)) =(u^{\rm eq}_{\mathbf k}, v_{\mathbf k}^{\rm eq})$ and study its
evolution in the presence of the periodic drive till a time $t_f$
which correspond to $N$ cycles of the drive $t_f =NT=2\pi N/\omega$,
where $N$ is an integer.

In order to study such dynamics, we focus on the following key
observables. First, we define the wavefunction of the BCS system 
$|\psi(t)\rangle=\prod_{\mathbf k}|\psi_{\mathbf k}(t)\rangle$, where
we have denoted
\begin{eqnarray}
|\psi(t)\rangle &=& \prod_{\mathbf k} \left(u_{\mathbf k}(t) + v_{\mathbf k}(t)
\hat{c}_{\mathbf k}^{\dagger} \hat{c}_{{\bf -k}}^{\dagger}\right) |0\rangle,
\label{neqwav}
\end{eqnarray}
with $u_{\mathbf k}(t)$ and $v_{\mathbf k}(t)$ being solutions of  \
\eref{bdg2} and \eref{tdsc1}. We now compute the effective magnetization $m(t)$,
defined as
\begin{eqnarray}
m_{\mathbf k}(t) &=& \langle \psi_{\mathbf k}(t)| \tau_z |\psi_{\mathbf k}(t)\rangle, \nonumber \\
m(t) &=& \sum_{\mathbf k} m_{\mathbf k}(t) = \sum_{\mathbf k} [1-2|v_{\bf
k}(t)|^2 ],\label{eq:mag}
\end{eqnarray}
where $\tau_z$ is the Pauli matrix in particle-hole space. The
observable $m(t)$ shall be equal to $m(t_i)$ after a full drive
cycle at $T=2\pi/\omega$ both in the impulse (where the wavefunction
does not have time to adjust to the drive) and adiabatic limit
(where the system remains in the ground state of the instantaneous
Hamiltonian). Note that $m(t)$ is the magnetization of the Ising or
Kitaev models described by BdG-like equations in their fermionic
representations sans the self-consistency
condition\ \cite{diptiman,arnab1}. For BCS fermions, $m(t)$ can be
easily related to the time-dependent fermion density $n(t)$
using the relation
\begin{eqnarray}
n(t) &=& \sum_{{\mathbf k}} \langle \psi_{\bf k}(t)|\left(\sum_{\sigma}\hat{c}^{\dagger}_{\bf k \sigma}\hat{c}^{ }_{\bf k \sigma}\right)|\psi_{\bf k}(t)\rangle = 2 \sum_{\mathbf k}
|v_{\mathbf k}(t)|^2 = 1-m(t), \label{mdef}
\end{eqnarray}
Thus $m(t)$ proves to be useful in comparing the behaviour of
integrable Ising and Kitaev models with that of the non-integrable
self-consistent BCS model. To this end, we also define the long time
average of $m(t)$
\begin{equation}
\label{eq:mt} Q \equiv \lim \limits_{n \to \infty}
\frac{1}{nT}\int^{nT}_0 {\mathrm d}t \times m(t),
\end{equation}
which shall also be used for such comparisons.

The second quantity which we compute is the wavefunction overlap. To
compute this, we first define the amplitudes $u_{\mathbf k}^{\rm ad}(t)$
and $v_{\mathbf k}^{\rm ad}(t)$ which correspond to the values of
$u_{\mathbf k}$ and $v_{\mathbf k}$ at time $t$ for adiabatic evolution. 
The amplitude $u_{\mathbf k}^{\rm ad}(t)$ is always real in our choice of gauge.
The ground state of $H$ with $\mu=\mu(t)$ can be written in terms of
these quantities as
\begin{eqnarray}
|\psi^{\rm ad}(t)\rangle &=& \prod_{\mathbf k} \left(u_{\mathbf k}^{\rm
ad}(t) + v_{\mathbf k}^{\rm ad}(t) \hat{c}_{\mathbf k}^{\dagger} \hat{c}_{{\bf
-k}}^{\dagger}\right) |0\rangle,
\nonumber\\
u^{\rm ad}_{\mathbf k}(t)  &=& \frac{1}{\sqrt
2} \bigg\{ 1+ \frac{\left[\epsilon_{\mathbf k} -\mu(t)\right]}{E({\mathbf k};t)}
\bigg\}^{1/2},\nonumber\\
v^{\rm ad}_{\mathbf k}(t) &=& \frac{1}{\sqrt
2}\bigg\{ 1- \frac{\left[\epsilon_{\mathbf k} -\mu(t)\right]}{E({\mathbf k};t)}
\bigg\}^{1/2} e^{-i\phi_0(t)}.\label{eqsolad}
\end{eqnarray}
where $E({\mathbf k};t)=\sqrt{[\epsilon_{\mathbf k} -\mu(t)]^2
+|\Delta(t)|^2}$, and $|\psi^{\rm ad}(t)\rangle$ is the adiabatic ground state continued in time. Also,
$\phi_0(t)$ is the phase of $\Delta(t)$. Note that $|\psi^{\rm ad}(t_f)\rangle= |\psi^{\rm
ad}(t_i)\rangle$ at the end of any integer number of drive cycles.
We now define the wavefunction overlap $F$
as
\begin{eqnarray}
F &=& |\langle \psi^{\rm ad}(t)|\psi(t)\rangle|^2 \nonumber\\
&=& \prod_{\mathbf k} F_{\mathbf k} = \prod_{\mathbf k} |u^{\rm ad  }_{\bf
k}(t) u_{\mathbf k}(t) + v^{\rm ad  }_{\mathbf k}(t) v_{\mathbf k}(t)|^2.
\label{wavover1}
\end{eqnarray}
The defect density or the density of excitations generated due the
dynamics at any instant of time can be written in terms of $F$ as
\begin{eqnarray}
\rho_d &=& \sum_{\mathbf k} \rho_d({\mathbf k}) \\
\rho_d({\mathbf k}) &=& 1-F_{\mathbf k} =  |u^{\rm ad  }_{\mathbf k}(t)
v_{\mathbf k}(t) - v^{\rm ad  }_{\mathbf k}(t) u_{\mathbf k}(t)|^2. \nonumber
\label{defect1}
\end{eqnarray}
Note that the defect density identically vanishes for adiabatic
dynamics and thus provides a suitable measure for deviation from
adiabaticity.

\begin{figure}[h!bt]
\rotatebox{0}{\includegraphics*[width=\linewidth]{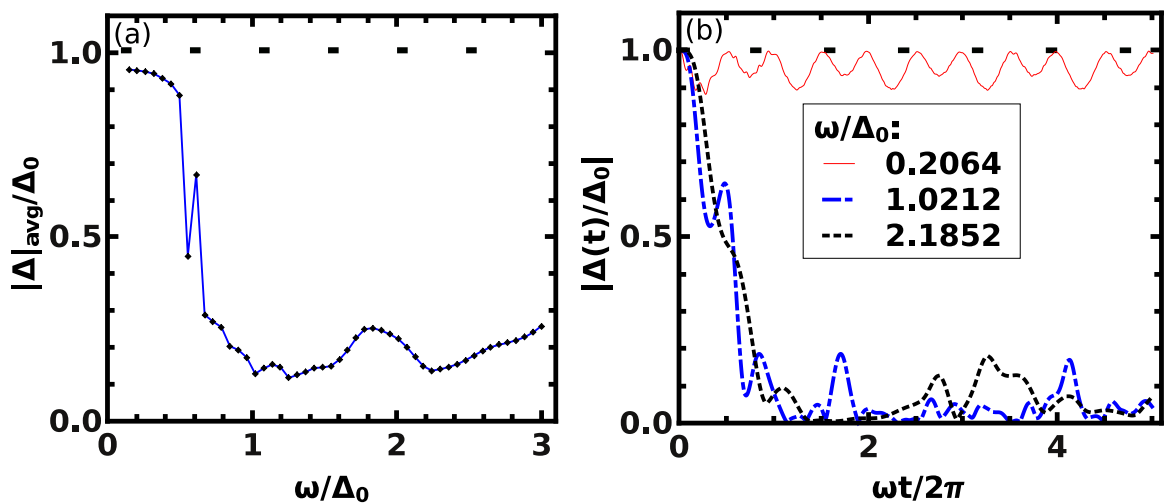}}
\caption{(Colour online) Plots of the long time average and
instantaneous values of the amplitude of the order parameter
$\Delta(t)$. The left panel shows the plot of the long time averaged
$|\Delta(t)|$ (averaged over $10$ cycles of the drive) with respect
to the drive frequency $\omega$. The right panel shows variation of
the instantaneous values of $|\Delta(t)|$ with time $t$ (in units of
$2\pi/\omega$) for several representative values of $\omega$
indicated in the legend. In both panels, $\Delta_0$ is indicated by
a black dashed horizontal line.} \label{fig1}
\end{figure}
Finally, we shall compute the residual energy which is the
additional energy put in the system due to the drive. This is
defined as the difference between the energy of the system at time
$t$ and the adiabatic ground state energy and can be written as
\begin{eqnarray}
E_r (t) &=& \sum_{\bf k } \left[\langle \psi_{\mathbf k}(t)|h_{\bf
k}(t)|\psi_{\mathbf k}(t)\rangle \right. \nonumber\\
& &\left. - \langle \psi_{\mathbf k}^{\rm ad}(t)
|h_{\mathbf k}(t)|\psi_{\mathbf k}^{\rm ad}(t)\rangle\right]\nonumber \\
&=& \frac{1}{g}\left[\Delta^2(t)+\Delta^{\ast
2}(t)-2\Delta^2_0\right] \nonumber\\
& &-\sum_{\mathbf k} \left[\epsilon_{\mathbf k} - \mu(t) \right][m_{\bf
k}(t)- m^{\rm ad}_{\mathbf k}(t)],
\end{eqnarray}
where $m_{\mathbf k}^{\rm ad} = 1-2 |v^{\rm ad}_{\mathbf k}|^2$. Note that
the residual energy also vanishes for adiabatic dynamics.

Before closing this section, we note that the BCS self-consistency
condition imparts dynamics to the order parameter $\Delta$ which can
be written, using  \ \eref{bdg2} and \eref{tdsc1}, as
\begin{eqnarray}
\dot{\Delta}(t) &=& i g \Big \{\Delta^\ast(t)m(t) + 2 \sum_{\mathbf k}
\left[\epsilon_{\mathbf k}-\mu(t) \right] u^\ast_{\mathbf k}(t) v_{\mathbf k}(t)
\Big \}. \label{deltaeq1} \nonumber\\
\end{eqnarray}
If the time dependence of $\Delta$ is ignored or rendered
negligible, then the system reverts to an ensemble of decoupled
two-level systems in momentum space with constant gap $\Delta_0$. In
this case, the dynamics is described by Landau-Zener-St\"uckelberg
theory~\cite{review:lzstls}. As we shall see in the next section, we
reach this regime for $ \omega/\Delta_0 \ll 1$; however, the
behaviour of a BCS system differs significantly from that of a bunch
of decoupled two-level system for moderate $\omega$ for which
$\omega/\Delta_0 \sim 1$.

\section{Numerical results}
\label{numerics}

In this section, we discuss the self-consistent numerical evaluation
of  \ \eref{bdg2} and \eref{tdsc1} for $d=2$ and subsequent
computations of $m(t)$, $Q$, $|\Delta(t)|$, $\rho_d$, and $E_r$ as
defined in\ \sref{formalism}. We have solved  \ \eref{bdg2} and
\eref{tdsc1} for BCS fermions in a $144\times144$ square optical
lattice and having  unit hopping amplitude ($J=1$). The equilibrium
gap and chemical potential has been taken to be $\Delta_0=0.1$ and
$\mu_0=0.01$ respectively. The periodic drive term has been taken to
be of the form $\mu_a \sin(\omega t)$ with $\mu_a=0.1$.

\begin{figure}
\rotatebox{0}{\includegraphics*[width=\linewidth]{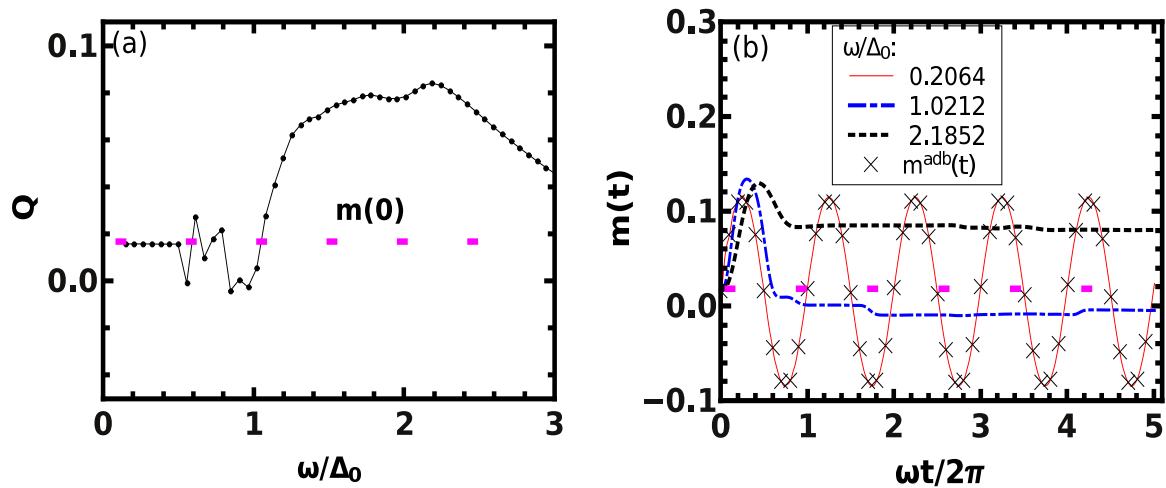}}
\caption{(Colour online) Left Panel: Plot of $Q$ as a function of
$\omega$ with averaging carried over $10$ drive cycles. Right panel:
Plot of $m(t)$ as a function of $\omega t/2\pi$ for representative
values of $\omega$ indicated in the legend. A few representative
values of the adiabatic magnetization $m^{adb}(t)$ is shown using
crosses. In both panels, the initial values of $Q$ and $m$ is
indicated by a magenta dashed horizontal line and all parameters are
same as in\ \fref{fig1}.} \label{fig2}
\end{figure}

We first consider the plot of the gap amplitude $|\Delta|$ in\ \fref{fig1}. 
The left panel of\ \fref{fig1} shows the average gap
amplitude as a function of the drive frequency $\omega$ after an
average over $10$ cycles. We find that the average value of the gap
amplitude decreases rapidly with increasing frequency and keeps
fluctuating about $|\Delta| \simeq 0.2 \Delta_0$ for large
$\omega/\Delta_0 \ge 2$. The right panel shows a plot of
$|\Delta(t)|/\Delta_0$ as a function of $\omega t/2\pi$. We find
that at small $\omega \ll \Delta_0$, $|\Delta(t)|$ displays
oscillatory behaviour with maximum and minimal values of $\Delta_0$
and $0.9 \Delta_0$ respectively. However, for $\omega \ge \Delta_0$,
the behaviour of $|\Delta(t)|$ is qualitatively different; it
decreases rapidly to near-zero values within the first couple of
drive cycles ($\omega t/2\pi \le 2$) and continues to fluctuate
around this value for longer drive times, never returning close to
its original value $\Delta(0)$. We note that such a behaviour of
$|\Delta(t)|$ clearly reflects the importance of the
self-consistency condition in the dynamics; any analysis with
$|\Delta(t)| \simeq \Delta_0$ at all times is expected to produce
qualitatively wrong results for $\omega \ge \Delta_0$.

\begin{figure}
\rotatebox{0}{\includegraphics*[width=\linewidth]{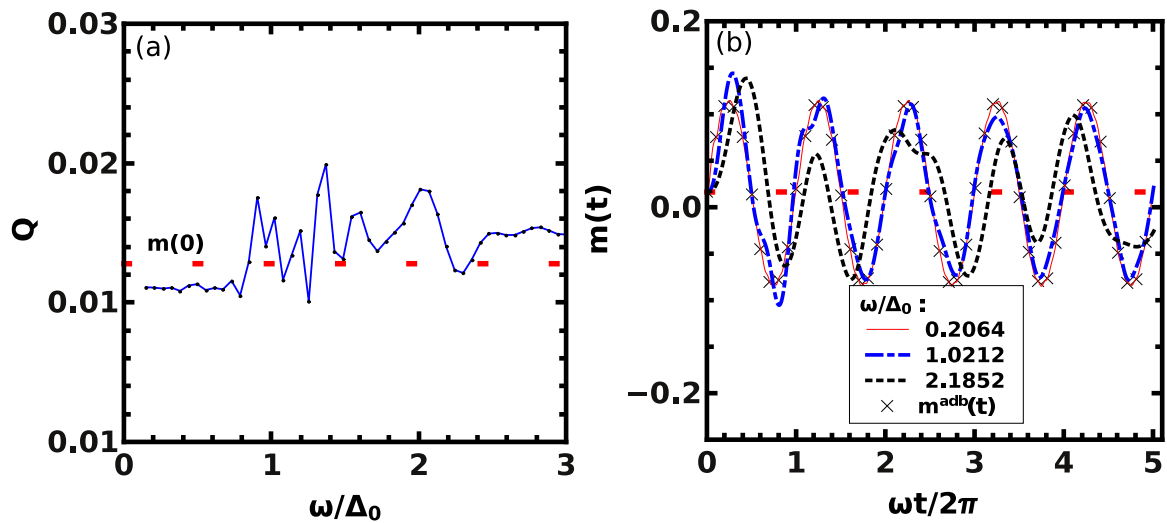}}
\caption{(Colour online) Same as in \ \fref{fig2} but for the
non-self-consistent dynamics.} \label{fig3}
\end{figure}

Next, we plot the effective magnetization $m(t)$ as a function of
time $t$ and its time average $Q$ as a function of the drive
frequency $\omega$. For the {\it self-consistent} dynamics
appropriate for fermions in the BCS regime, as shown in \
\fref{fig2}, there are clearly three regimes, one crossing over to
the other as $\omega$ is increased. For $\omega \ll \Delta_{0}$,
$m(t)$ (right panel) oscillates with large amplitude, following the
drive almost adiabatically, resulting in $Q = m(0)$. The
oscillations, though large, respects the symmetry of the drive, \textit{i.e.},
the long time average of the magnetization vanishes with the DC part of the 
drive, \textit{viz.} the equilibrium chemical potential $\mu_0$. As
$\omega$ approaches $\Delta_{0}$, this symmetry is destroyed,
resulting $Q \ne m(0)$. For $\omega > \Delta_{0}$, the oscillatory
behaviour of $m(t)$ with large amplitude is replaced by relaxation to
an approximately constant value (with negligible fluctuations)
within few initial cycles (right panel). This constant value
determines the value of $Q$ (left panel), and we find that it
deviates steadily from $m_{0}$ as $\omega$ is increased for $\omega
< 2.5 \Delta_0$. We note that for $\omega > \Delta_0$, the mixing of
the ${\mathbf k}$ modes of the quasiparticle excitations, which
originates from the presence of the self-consistency condition and
is therefore absent in Ising or Kitaev systems, is at the heart of
such a deviation. Any hysteresis or freezing of the magnetization that would cause the 
drive symmetry to break was observed in periodically driven
transverse Ising chains for large amplitudes and frequencies\ \cite{arnab1}, and is also seen 
here for the self-consistent case at $\omega \lesssim \Delta_0$. For very
large $\omega$, we of course see the behaviour of $m(t)$ crossing
over to a regime where $Q \rightarrow m(0)$ again -- here $\omega$
becomes too large for the system to react at all, and $m(t)$ remains
frozen around $m(0)$ for all time (the regime sets in beyond $\omega
> 2.4 \mbox{ } \Delta_0$).

The above behaviour is to be contrasted with the {\it
non-self-consistent} case summarized in\ \fref{fig3}. Here
$m(t)$ always executes a large, almost synchronized oscillation,
approximately following the adiabatic path (black crosses in the
right panel of\ \fref{fig3}). Naturally, the resulting values of
$Q$ are close to $m(0)$ (albeit with some small fluctuations). This
suggests that the synchronous oscillation is simply a manifestation
of the near-adiabatic nature of the dynamics. Synchronization could
also occur due to self dephasing of the system, after all the
transients (some of them having power-law tails) have died down, due
to quantum interference between the modes \cite{arnab1,AD-Gen-Pic}.
But such synchronization would appear only in the $\omega t
\rightarrow \infty$ limit, unlike in the present case, where the
effect is visible from the very first cycle. The qualitative
departure from this behaviour in the self-consistent case seems to
stem from the non-adiabaticity induced by the self-consistency
condition  \eref{tdsc1} which makes the effective Hamiltonian
non-linear. The overlapping eigenfunctions of the non-linear
Hamiltonian makes the criteria for adiabatic behaviour much more
restricted compared to a linear case (see. e.g., Yukalov\
\cite{Yukalov}). We shall address the behaviour of $m_{{\mathbf k}_F}(t)$
in a more quantitative manner in\ \sref{analytic}, where we
shall show that the value of $m_{{\mathbf k}_F}$ after one drive cycle decays
as $1/\omega$ for $\omega \gg \Delta_0$.

\begin{figure}
\rotatebox{0}{\includegraphics*[width=\linewidth]{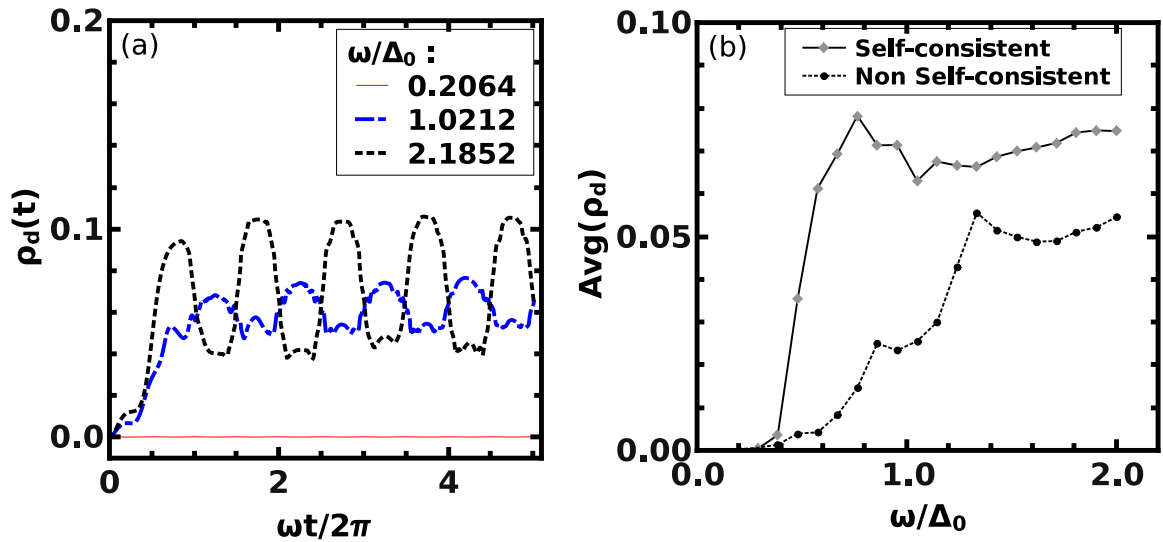}}
\caption{(Colour online) Left panel: Plot of the instantaneous defect
density $\rho_d(t)$ as a function of $\omega t/(2\pi)$. The values
of the drive frequency $\omega$ is indicated in the inset. Right
panel: Plots of the long time average (averaged over $10$ drive
cycles) of the defect density (both the non-self-consistent and the
self-consistent cases) as a function of the drive frequency
$\omega$.} \label{fig4}
\end{figure}

\begin{figure}
\rotatebox{0}{\includegraphics*[width=\linewidth]{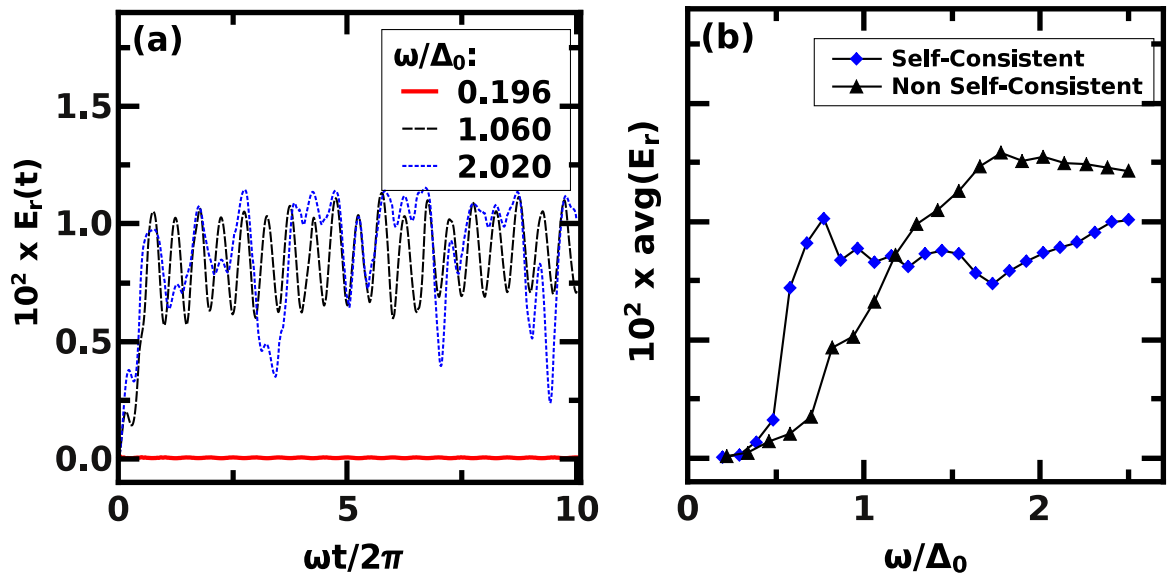}}
\caption{(Colour online) Same as in\ \fref{fig4} but for residual
energy $E_r$.} \label{fig5}
\end{figure}

\begin{figure}
\rotatebox{0}{\includegraphics*[scale=0.6]{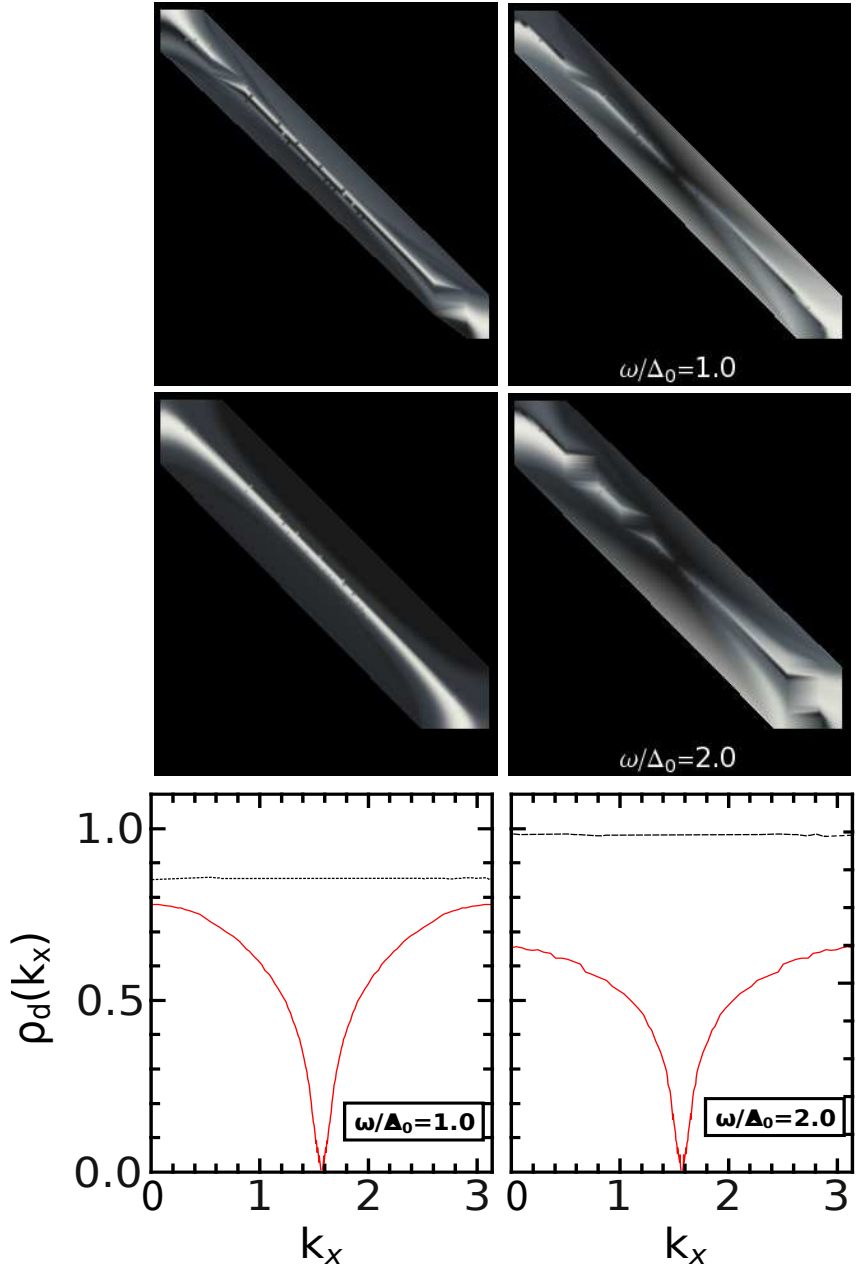}}
\caption{(Colour online) Top (Middle) Panels: Plot of the momentum
distribution of the defect density $\rho_d({\mathbf k})$ (where $\mathbf{k}=k_x\hat{x}+k_y\hat{y}$) as a function 
of $k_x$ (abscissas) and $k_y$ (ordinates) after a full
drive cycle for s-wave [left panels] and d-wave [right panels]
pairing symmetries. The drive frequencies are specified in the inset
of the right column. In these plots, lighter shades represent higher
quasiparticle excitation densities. The abscissas of the top and middle panels have the same range 
and scale as those of the bottom panels \textit{viz.} $k_x\in[0,\pi]$, and have been omitted for brevity. 
The ordinates of the top and middle panels plot $k_y\in[0,\pi]$ in a similar manner, and have been omitted as well.\\
Bottom Panel: Plot of the momentum distribution of the defect density (black dotted line for
s-wave and red solid line for d-wave) along the Fermi surface in the
first quadrant ($k_x,k_y \ge 0$) as a function of $k_x$, with $k_y$ chosen to 
lie on the Fermi surface given by $f_{\bf k}=\cos{k_x}+\cos{k_y}-\mu_0=0$ $\forall k_x$. These plots
clearly demonstrate the dip in the momentum distribution at
the node for d-wave pairing. Analogous dips occur at other three
quadrants at the position of the other nodes of $\Delta({\mathbf k})$. In all of these plots, $\mu_0=0.01$, $\Delta_0=0.1$.}
\label{fig6}
\end{figure}

Next, we consider the behaviour of the defect density $\rho_d$ as
shown in\ \fref{fig4}. Here, as expected, the defect density
becomes significant only for $\omega \ge \Delta_0$. The plot of the
self-consistent defect dynamics shown in the left panel of\
\fref{fig4} demonstrates that the defect density is an oscillatory
function of $\omega$. The time-averaged defect density shown in the
right panel of\ \fref{fig4} for both the non-self-consistent and
the self-consistent dynamics shows that these quantities display
qualitatively similar behaviour. A similar behaviour is seen for the
residual energies as can be seen from\ \fref{fig5}. We find that
$E_r(t)$ vanishes for $\omega \ll \Delta_0$ and displays oscillatory
behaviour for $\omega \ge \Delta_0$. The behaviour of residual energy
and the defect density clearly shows that the system wavefunction
never comes back to itself for any $\omega$; thus BCS superfluids do
not seem to support dynamic freezing as predicted for superfluid
bosons by Mondal, Pekker and Sengupta \ \cite{pekker1}.
Finally, we consider the momentum distribution of the defect density
at the end of a drive cycle and compare such plots for d-wave and
s-wave pairing symmetries. The generalization of our calculation for
d-wave pairing symmetry is straightforward and constitutes changing
$\Delta_0 \to \Delta_{\mathbf k} = \Delta_0 [\cos(k_x)-\cos(k_y)]$ in
 \ \eref{bdg2} and \eref{tdsc1}. The rest of the computation
follows exactly as charted out in\ \sref{formalism}. We expect
the momentum distribution of the defect density to be qualitatively
different for s- and d-wave pairing symmetries. For s-wave, the
defect density has a uniform pattern in the Brillouin zone as
expected from the momentum independence of the order parameter. In
contrast, for the d-wave pairing symmetry, $\Delta({\mathbf k})$
vanishes at $k_x=\pm k_y$. It is easy to see from  \
\eref{bdg2} that for such momenta, one has
\begin{eqnarray}
u_{\mathbf k}(t) &=& \theta(-f_{\mathbf k}) e^{ - i \int^t_{t_i}
[\epsilon_{\mathbf k} -\mu(t')] dt'}, \nonumber\\
v_{\mathbf k}(t) &=& \theta(f_{\mathbf k}) e^{ i \int^t_{t_i}
[\epsilon_{\mathbf k} -\mu(t')] dt' },
\end{eqnarray}
where $f_{\bf k}=\epsilon_{\bf k}-\mu_0$, and $u_{\mathbf k}(t_i) =\theta(-f_{\mathbf k})$ , $v_{\mathbf k}(t_i)=
\theta(f_{\mathbf k})$ are obtained by solving the BCS equations\
 \eref{bdg1} for $\Delta({\mathbf k})=0$. We note that this also
implies that at the end of a cycle, at $t=t_f= \omega/2\pi$ where
$\mu(t_f)=\mu(t_i)$, the phase integrals vanish and one obtains
$u_{\mathbf k}(t_f) = \theta(-f_{\mathbf k})$ and $v_{\mathbf k}(t_f) =
\theta(f_{\mathbf k})$. Thus the wavefunction overlap for
$\Delta({\mathbf k})=0$ at the end of a drive cycle becomes unity
leading to vanishing defect density at the nodes. However, away from
the nodes, where the $\Delta({\mathbf k})$ is finite, we expect high
density of quasiparticle excitations. This qualitative consideration
matches well with the numerical results shown in the right panels of
\ \fref{fig6}. In contrast, the s-wave pairing symmetry has a
constant $\Delta({\mathbf k}) =\Delta_0$ and hence leads to a uniform
defect density pattern as shown in left panels of \ \fref{fig6}.
This results in a qualitative difference between the defect density
patterns originating from superfluids with the two pairing
symmetries. We note that although we have explicitly calculated the
defect density for s- and $d_{x^2-y^2}$-wave symmetries in the
present work, the approach can be straightforwardly generalized to
other pairing symmetries. In general, we expect the defect density
to display a minimum at the position of the node of the gap. Thus
the momentum distribution of the defect density of a Fermi
superfluid bears the signature of the positions of the nodes of the
order parameters on the Fermi surface and hence can be used to
distinguish between various order parameter symmetries.

\section{Analytical computation of the magnetization}
\label{analytic}

In this section, we obtain an analytical understanding of the
behaviour of the magnetization or fermion density at the gap edge,
{\it i.e.} at $f_{\mathbf k}=0$, after a drive cycle in the
high-frequency limit. We note that from Refs.\
\cite{review:lzstls} and \cite{wittig:lzformula}, we can
conclude that the non self-consistent dynamics for a two-level
system described in\ \sref{formalism} is affected by two
phenomena, Landau-Zener tunnelling and the St\"uckelberg phase. In
what follows, we derive an analogous picture for the s-wave BCS
fermions {\it with the self-consistency condition}. The calculation
is carried out here for s-wave superfluids but can be easily
generalized to other pairing symmetries.

We begin with  \ \eref{bdg2} and\ \eref{tdsc1}, yielding
\begin{eqnarray}
\label{eq:selfconsistent}
\dot{u}_{\mathbf k}(t) &=& -i \left[\epsilon_{\mathbf k}-\mu(t) \right]u_{\mathbf k}(t) - i \Delta(t) v_{\mathbf k}(t),\nonumber \\
\dot{v}_{\mathbf k}(t) &=& i \left[\epsilon_{\mathbf k}-\mu(t) \right]v_{\mathbf k}(t) - i \Delta^\ast(t) u_{\mathbf k}(t),\nonumber \\
\Delta(t) &=& g\sum_{{\mathbf k}} u^\ast_{\mathbf k}(t)v_{\mathbf k}(t).
\end{eqnarray}
\begin{table*}
\caption{\label{tab:icsadb} Values of system dynamical variables
$u_{\mathbf k}(t)$, $v_{\mathbf k}(t)$, $s_{\mathbf k}(t)$, $b_{\mathbf k}(t)$, and
$\Delta(t)$, their time derivatives  and second derivatives at $t=0-$
where the avoided crossing takes place for adiabatic initial
conditions. See  \ \eref{eqsol},\ \eref{bdg2},\ \eref{tdsc1} and\
\eref{deltaeq1} for details.}
\begin{indented}
\item[]\begin{tabular}{@{}lllll}
        \br
    Value                          & &First derivative                                        & &Second derivative                                                              \\ \hline \hline
                    &                                          &           \\ 
         $u_{\mathbf k}=u^{\rm eq}_{\mathbf k}$ & &$\dot{u}_{\mathbf k}=-i\left(f_{\mathbf k} u^{\rm eq}_{\mathbf k}+\Delta_0 v^{\rm eq}_{\mathbf k}\right)$
         & &$\ddot{u}_{\mathbf k}=-\left[E^2({\mathbf k})-i\mu_a\omega\right]{u^{\rm eq}_{\mathbf k}}$                                     \\
                    &                                                  &         \\ 
                    &                                      &          \\
         $v_{\mathbf k}=v^{\rm eq}_{\mathbf k}$ & &$\dot{v}_{\mathbf k}=+i\left(f_{\mathbf k} v^{\rm eq}_{\mathbf k}-\Delta_0 u^{\rm eq}_{\mathbf k}\right)$
         & &$\ddot{v}_{\mathbf k}=-\left[E^2({\mathbf k})+i\mu_a\omega\right]{v^{\rm eq}_{\mathbf k}}$                                                             \\
                    &                                      &          \\ 
                    &                                  &        \\
    $s_{\mathbf k}=1$           & &$\dot{s}_{\mathbf k}=-if_{\mathbf k}$                    & &$\ddot{s}_{\mathbf k}=-\left(f^2_{\mathbf k}-i\mu_a\omega\right)$ \\
                    &                                  &        \\ 
                    &                                  &        \\
    $b_{\mathbf k}=v^{\rm eq}_{\mathbf k}$  & &$\dot{b}_{\mathbf k}=-i\Delta_0 u^{\rm eq}_{\mathbf k}$          & &$\ddot{b}_{\mathbf k}= -\Delta_0v^{\rm eq}_{\mathbf k}\left(\Delta_0+2f_{\mathbf k}\right)$ \\
                    &                                  &        \\ 
                    &                                      &          \\
         $\Delta=\Delta_0$      & &$\dot{\Delta}=0$                         & &$\ddot{\Delta}=-2i\Delta_0 \mu_a\omega$    \\
                    &                                      &          \\ \br
\end{tabular}
\end{indented}
\end{table*}
Defining the terms
\begin{eqnarray}
\label{eq:akbk}
s_{\mathbf k}(t) &=& \exp\left\{-i\int^t_0\mathrm{d}t' {\left[\epsilon_{\mathbf k}-\mu(t') \right]}\right\},\nonumber \\
u_{\mathbf k}(t) &=& a_{\mathbf k}(t) s_{\mathbf k}(t),\quad v_{\mathbf k}(t) =
b_{\mathbf k}(t) s^{-1}_{\mathbf k}(t),
\end{eqnarray}
the dynamics of the system can be rewritten as
\begin{eqnarray}
\label{eq:akbkdyn0}
\dot{a}_{\mathbf k} &=& -i\Delta(t)b_{\mathbf k}(t) s^{-2}_{\mathbf k}(t),\nonumber \\
\dot{b}_{\mathbf k} &=& -i\Delta^\ast(t)a_{\mathbf k}(t) s^{2}_{\mathbf k}(t),
\end{eqnarray}
which leads to two decoupled second-order differential equations for
$a_{\mathbf k}(t)$ and $b_{\mathbf k}(t)$ given by
\begin{eqnarray}
\label{eq:akbkdyn} \ddot{a}_{\mathbf k} -\left\{ 2i \left[\epsilon_{\bf
k}-\mu(t) \right]+\frac{\dot{\Delta}(t)}{\Delta(t)}\right\}
\dot{a}_{\mathbf k}
+ |\Delta(t)|^2 a_{\mathbf k} &=&0,\nonumber \\
\ddot{b}_{\mathbf k} +\left\{2i \left[\epsilon_{\mathbf k}-\mu(t)
\right]-\frac{\dot{\Delta}^\ast(t)}{\Delta^\ast(t)}\right\}
\dot{b}_{\mathbf k} + |\Delta(t)|^2 b_{\mathbf k} &=&0. \nonumber\\
\end{eqnarray}
The self-consistency condition can be written in terms of ${a}_{\bf
k}(t)$ and $b_{\mathbf k}(t)$ as
\begin{equation}
\label{eq:gap:akbk} \Delta(t) = g \exp{\left[-2i\int^t_0{\mathrm
d}t'\mu'(t')\right]} \sum_{\mathbf k}a^\ast_{\mathbf k}(t)b_{\mathbf k}(t)
e^{2if_{\mathbf k}t},
\end{equation}
where $f_{\mathbf k}= \epsilon_{\mathbf k}-\mu_0$ and
$\mu'(t)=\mu_a\sin{\omega t}$. The initial conditions for $a_{\bf
k}$ and $b_{\mathbf k}$ can be easily obtained from those of $u_{\mathbf k}$
and $v_{\mathbf k}$ as discussed in\ \sref{formalism}.

To obtain an analytical insight into the solution of these
equations, we note that there is an avoided crossing at $t_{1{\bf
k}} = t_1 = \arcsin(f_{\mathbf k}/\mu_a)/\omega$ and that $t_1$
approaches zero for large $\omega$ and on the Fermi surface. Further if $\omega t_1 \ll 1$, a condition which is exactly
satisfied at $f_{\mathbf k}=0$, we may use the Zener approximation
$\mu'(t)\approx\mu_a\omega t$ for $\mu(t)$ close to $t=t_1$ when the
system traverses the avoided
crossing~\cite{wittig:lzformula,landau:lzformula,zener:lzformula},
and restrict ourselves within the adiabatic impulse model where all
excitations away from the avoided crossing are
ignored~\cite{review:lzstls}. From the definitions in  \
\eref{eq:akbk} and  \ \eref{eq:selfconsistent}, we can simplify  \
\eref{eq:akbkdyn0} to yield
\begin{equation}
\label{eq:akdyn} \dot{a}_{\mathbf k} = -i\Delta(t) b_{\mathbf k}(t)
e^{2i\left(f_{\mathbf k}t-\frac{1}{2}\mu_a\omega t^2\right)}.
\end{equation}
We now assume that $\omega\gg\Delta_0$ and define
\begin{eqnarray}
x_{\mathbf k} &=& t \sqrt{\mu_a\omega} - f_{\mathbf k}/\sqrt{\mu_a\omega}
\nonumber\\
\theta_{\mathbf k}(t)&=&  b_{\mathbf k}(t) \Delta(t)/\Delta_0=v_{\bf
k}(t)s_{\mathbf k}(t)\Delta(t)/\Delta_0. \label{defeq}
\end{eqnarray}
We can use these definitions to simplify  \ \eref{eq:akdyn},
yielding
\begin{equation}
\frac{\partial a_{\mathbf k}}{\partial x_{\mathbf k}} = -i
\frac{\Delta_0\theta_{\mathbf k}(x_{\mathbf k})}{\sqrt{\mu_a\omega}}
e^{\frac{if^2_{\mathbf k}}{\mu_a\omega}} e^{-ix^2_{\mathbf k}}.
\end{equation}
Thus, the amplitude $a_{\mathbf k}(t)$ after the system traverses an
avoided crossing is approximately given by
\begin{equation}
\label{eq:afk} \mathcal{A}^{(1)}_{\mathbf k} - \mathcal{A}^{(0)}_{\mathbf k}
= - \frac{i\Delta_0}{\sqrt{\mu_a\omega}}e^{\frac{if^2_{\bf
k}}{\mu_a\omega}} \int^\infty_{-\infty} \mathrm{d} x_{\mathbf k}
\theta_{\mathbf k}(x_{\mathbf k}) e^{-ix^2_{\mathbf k}},
\end{equation}
where $\mathcal{A}^{(n)}_{\mathbf k}$ denotes the amplitude $a_{\mathbf k}$
after $N$ passages across the avoided crossings with
$\mathcal{A}^{(0)}_{\mathbf k}$ being the initial amplitude of $a_{\bf
k}$\ \cite{wittig:lzformula,review:lzstls}.

The integral in the right side of  \ \eref{eq:afk} can be evaluated
by contour integration whose details are charted out in the
Appendix. This yields
\begin{eqnarray}
\label{eq:afkfinal} \mathcal{A}^{(1)}_{\mathbf k} &=& u^{\rm eq}_{\bf
k}- \sqrt{\frac{\pi\Delta^2_0}{\mu_a\omega}}
e^{i\left(\frac{f^2_{\bf
k}}{\mu_a\omega}+\frac{\pi}{4}\right)} \nonumber\\
& &\times \sum^\infty_{n=0} \frac{1}{n!}
\frac{\left(-i\right)^n}{\left(4\mu_a\omega\right)^n}
\frac{\partial^{2n}\theta_{\mathbf k}}{\partial^{2n}t}\bigg|_{t=t_{\bf
k}},
\end{eqnarray}
where we have used $a_{\mathbf k}(0)=u^{\rm eq}_{\mathbf k}$ from   \
\eref{eq:akbk}, and $t_{\mathbf k}= f_{\mathbf k}/(\mu_a \omega)$. Note that,
in general, $t_{\mathbf k}\neq 0$ and so evaluating the modified Landau
Zener probability for an arbitrary momentum will require knowledge
of the system at times $t_{\mathbf k}$. However, $t_{\mathbf k}$ vanishes
exactly on the Fermi surface (characterized by $f_{\mathbf k}=0$) and
can be set to zero for all ${\mathbf k}$ that lie within
$\mathcal{O}(\mu_a\omega/v_F)$ around the Fermi surface, where $v_F$
is the Fermi velocity. In the rest of this section, we shall
restrict ourselves to this limit.

The fermion density in momentum space $n_{{\mathbf k}}= 2 |\mathcal{B}^{(1)}_{\mathbf k}|^2= 2
 |v_{\mathbf k}|^2$ after one passage across the avoided crossing can
be obtained in terms of the modified Landau Zener probability
\begin{eqnarray}
\label{eq:lzmod}
|\mathcal{B}^{(1)}_{\mathbf k}|^2 &=& 1 - |\mathcal{A}^{(1)}_{\mathbf k}|^2 \nonumber \\
    &=& 1-\Big[\left(u^{\rm eq}_{\mathbf k}\right)^2+\chi_0|c_{\mathbf k}|^2 \nonumber\\
& &- 2u^{\rm eq}_{\mathbf k}\sqrt{\chi_0}\times{\rm Re}\left(c_{\bf
k}e^{i\kappa_{\mathbf k}}\right)\Big],
\end{eqnarray}
where we have defined
\begin{eqnarray}
\label{eq:ckchieta} c_{\mathbf k} &=& \sum^\infty_{n=0} \frac{1}{n!}
\frac{\left(-i\right)^n} {\left(4\mu_a\omega\right)^n}
\frac{\partial^{2n}\theta_{\bf
k}} {\partial^{2n}t}\bigg|_{t=t_{\mathbf k}}, \nonumber\\
\kappa_{\mathbf k} &=& \frac{f^2_{\mathbf k}}{\mu_a\omega}+\frac{\pi}{4},
\quad \chi_0 = \frac{\pi\Delta^2_0}{\mu_a\omega}.
\end{eqnarray}
{Noting that for $\omega \gg f_{\mathbf k}$, $\kappa_{\mathbf k} \sim
\pi/4$, and approximating}
\begin{figure}[h!bt]
\rotatebox{0}{\includegraphics*[width=\linewidth]{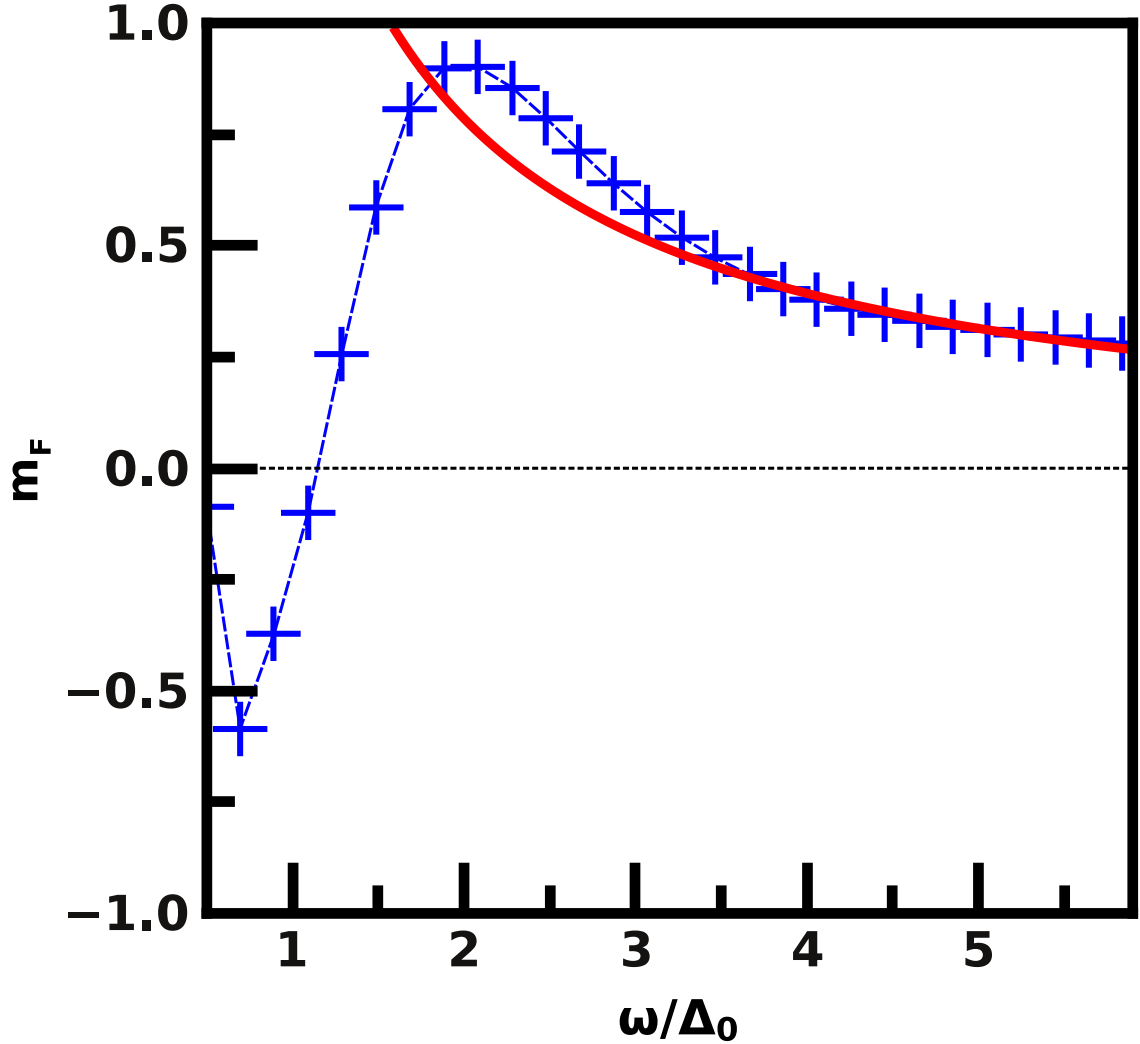}}
\caption{(Colour Online) Numerical plots (blue crosses) of $m_{F}$
as a function of $\omega$. The system parameters are the same as in
\ \fref{fig1}. The red solid line indicates the analytical result
obtained in  \ \eref{meqfinal}.} \label{fig:magcomp}
\end{figure}
\begin{equation}
\label{eq:ck0} c_{\mathbf k} \approx \sum^\infty_{n=0} \frac{1}{n!}
\frac{\left(-i\right)^n}{\left(4\mu_a\omega\right)^n}
\frac{\partial^{2n}\theta_{\mathbf k}}{\partial^{2n}t}\bigg|_{t=0},
\end{equation}
one finally gets an expression for the modified Landau-Zener
probability
\begin{eqnarray}
\label{eq:lz:pert} |\mathcal{B}^{(1)}_{\mathbf k}|^2 &=& \left(v^{\rm
eq}_{\mathbf k}\right)^2 -\chi_0|c_{\mathbf k}|^2+u^{\rm eq}_{\bf
k}\sqrt{2\chi_0}{\rm Re}\left[\left(1+i\right)c_{\mathbf k}\right]
\nonumber\\
\end{eqnarray}
Each of the terms in the sum of  \ \eref{eq:ck0} can be obtained
from\ \tref{tab:icsadb} and higher order derivatives thereof
using  \ \eref{deltaeq1} and either  \ \eref{eq:akbkdyn0} or  \
\eref{eq:akbkdyn} at $t=0$. We now define $P_{\mathbf k}$ to be the
traditional Landau Zener probability for the non self-consistent
case~\cite{landau:lzformula,zener:lzformula,wittig:lzformula}
\textit{viz.}
\begin{equation}
\label{lzsreg}
 P_{\mathbf k} = e^{-\chi_0}.
\end{equation}
Now, we can write $|\mathcal{B}^{(1)}_{{\mathbf k}}|^2  = (P_{\mathbf k}/2)
e^{\gamma_{\mathbf k}}$, where
\begin{eqnarray}
\label{eq:modlzs} \gamma_{\mathbf k} &=& \chi_0 + \ln\Big\{2\left(v^{\rm
eq}_{\mathbf k}\right)^2-2\chi_0|c_{\mathbf k}|^2
\nonumber\\
& &+2u^{\rm eq}_{\mathbf k}\sqrt{2\chi_0}{\rm
Re}\left[\left(1+i\right)c_{\mathbf k}\right]\Big\}
\end{eqnarray}
for large $\omega$. We now investigate regions close to the Fermi
surface by simplifying  \ \eref{eq:ck0}, retaining only the terms
up to $n=1$ in the expansion. This yields
\begin{equation}
 \label{eq:ck0pert}
c_{\mathbf k} \approx \frac{v^{\rm eq}_{\bf
k}}{2}\left[1+\frac{i\chi_0}{2\pi}\left(1+\frac{2f_{\bf
k}}{\Delta_0}\right)\right],
\end{equation}
where we have used expressions from\ \tref{tab:icsadb}. Taking
the approximation for $c_{\mathbf k}$ in eq\ \eref{eq:ck0pert},
substituting its value into  \ \eref{eq:lzmod}, and retaining
lowest contributing orders of $\chi_0$ yields
\begin{equation}
\label{eq:b1} |\mathcal{B}^{(1)}_{\mathbf k}|^2  \approx |v^{\rm
eq}_{\mathbf k}|^2\left[1+\frac{\chi^{1/2}_0}{\sqrt{2}}\frac{u^{\rm
eq}_{\mathbf k}}{v^{\rm eq}_{\mathbf k}}\right].
\end{equation}
The occupation amplitude $\mathcal{B}^{(1)}_{\mathbf k}$ is realized
after the first passage across the avoided crossing and when the
second passage begins. The passage starts when the adiabatic energy
equals the BCS gap $\Delta_0$. The adiabatic energies are given by
$E({\mathbf k};t)=\pm \sqrt{\left[f_{\mathbf k}-\mu_a\sin{\omega
t}\right]^2+|\Delta_0|^2}$, which is $E({\mathbf k})$ from\ \eref{eqsol},
with $\mu_0$ replaced by $\mu_0+\mu_a\sin{\omega t}$. The passage
ends when the velocity of the adiabatic energy vanishes,
\textit{i.e} when $\dot{E}({\mathbf k};t)=0$ at  $t=\pi/\omega$ or half
a period. Thus, $|\mathcal{B}^{(1)}_{\mathbf k}|^2$ is the fermion
density after half a drive cycle. After one complete period
\textit{i.e} two passages across the avoided crossing, the fermion
density in the adiabatic impulse limit is given
by~\cite{review:lzstls}
\begin{eqnarray}
\label{eq:prob:1p}
 |\mathcal{B}^{(2)}_{\mathbf k}|^2 &=& 4 |\mathcal{B}^{(1)}_{\mathbf k}|^2
 \left(1-|\mathcal{B}^{(1)}_{\mathbf k}|^2\right) \sin^2[\Phi_{\rm st}] \nonumber\\
 &\simeq&  2 |\mathcal{B}^{(1)}_{\mathbf k}|^2
 \left(1-|\mathcal{B}^{(1)}_{\mathbf k}|^2\right)\nonumber \\
 &=&2\bigg[\left(u^{\rm eq}_{\mathbf k}v^{\rm eq}_{\mathbf k}\right)^2\left(1-\frac{\chi_0}{2}\right) \nonumber \\
 & & +u^{\rm eq}_{\mathbf k}v^{\rm eq}_{\mathbf k}\left(|u^{\rm eq}_{\mathbf k}|^2-|v^{\rm eq}_{\mathbf k}|^2\right)\sqrt{\frac{\chi_0}{2}}\bigg],
\end{eqnarray}
where in the second line, we have used the fact that the
St\"uckelberg phase $\Phi_{\rm st} \to \pi/4$ for large $\omega$
\cite{review:lzstls}. Thus the magnetization $m_{\mathbf k}^{(2)}$ after
one period (or two passages across the avoided crossing) can be
evaluated using  \ \eref{eq:mag} and\ \eref{eq:prob:1p}, yielding
\begin{eqnarray}
\label{eq:m2:ex} m^{(2)}_{\mathbf k} &\simeq&
1-4\left(1-\frac{\chi_0}{2}\right) \left(u^{\rm eq}_{\mathbf k}v^{\rm
eq}_{\mathbf k}\right)^2 +\sqrt{8\chi_0} u^{\rm eq}_{\mathbf k}v^{\rm
eq}_{\mathbf k}m^{\rm eq}_{\mathbf k}, \nonumber\\
\end{eqnarray}
where $m^{\rm eq}_{\mathbf k}$, the equilibrium magnetization, is given
by $m^{\rm eq}_{\mathbf k}=1-2|v^{\rm eq}_{\mathbf k}|^2$. Thus, on the Fermi surface,
where all approximations used to arrive at this result are
clearly valid, one obtains, using $u_{F}^{\rm eq} (v_{F}^{\rm
eq})=u_{{\bf k}={\mathbf k}_F}^{\rm eq} (v_{{\bf k}={\mathbf k}_F}^{\rm
eq})={1\over\sqrt{2}}$,
\begin{eqnarray}
m_{F}\equiv m^{(2)}_{{\bf k}={\mathbf k}_F}  = m_0 \frac{\Delta_0}{\omega}, \quad
m_0 = \frac{\pi \Delta_0}{2 \mu_a}. \label{meqfinal}
\end{eqnarray}
Here, ${\mathbf k}_F$ denotes the momentum vector on the Fermi surface.
We note that $m_{F}$ does not depend on the orientation of ${\mathbf k}$ on
the Fermi surface. This is a consequence of the s-wave symmetry of
the superfluid order parameter and is not going to be present for
other pairing symmetries.

Thus, we find that the frequency dependence of the magnetization (or
equivalently the fermion density) is $m_{F} \sim\omega^{-1}$. The
magnetization drops off at the same manner as the non
self-consistent case (\textit{i.e} the driven Ising model where
$m\sim\omega^{-1}$ can be obtained from the Landau Zener formula).
The constant $m_0$ which decides the rate of the decrease of
$m_{F}$ with $\omega$, however, is different in the two cases. In
the non self-consistent case, $\Delta$ is always constant and so
$c_{\mathbf k}={1\over\sqrt{2}}$ exactly as yielded by  \ \eref{eq:ck0}.
Thus, $m^{(2)}\approx 2\chi_0$ to lowest order which is four times
its value for the self-consistent case. We note here that although
we have concentrated on $m_{F}$, our results are expected to be
accurate for all $m_{\mathbf k}$ for which $t_1 \simeq 0$ and $f_{\bf
k}\ll \mu_a$. The generalization of this treatment to $N$ periods
seems to be difficult due to the necessity of taking into account
multiple Stuckelberg phases and we leave this issue for a possible
future study.

To check the accuracy of the analytical result, we compare\ 
\eref{meqfinal} with numerical results for the magnetization on the
Fermi surface after one drive period for a $144\times 144$ square
optical lattice which is very close to half filling ($\mu_0=0.01$)
with $\mu_a=\Delta_0=0.1$. At each time step of the numerics, the
Fermi surface was strobed with a tolerance $\sim \mu_a \omega$. In
the limit of $f_{\mathbf k} \ll \mu_a$, the variation of $m_{F}$ is
expected to be small within this region of the Brillouin zone and an
average over all momenta inside this region should yield values
close to that predicted by  \ \eref{meqfinal}. The agreement, as
shown in \ \fref{fig:magcomp}, is quite good for $\omega \ge 3
\Delta_0$ but poor for smaller $\omega/\Delta_0$ where some of the
approximations made in this section are clearly violated.

\section{Discussion}
\label{discuss}
 Experimental verification of our work will require generation
of a time-dependent chemical potential. This can be easily done by
turning on an additional trap with oscillatory time dependence
leading to a potential of the form $\kappa(t) r^2/2$. Both the
confining and the additional trap potentials are to be made
wide-enough so that the atoms residing at the centre of the trap
feel an almost spatially constant chemical potential. Such traps can
be easily designed in current experimental setups \cite{comment1}.
To verify our theory, we propose momentum distribution measurements
as done recently for fermions on a honeycomb lattice by
Tarruell, Greif, Uehlinger, Jotzu and Esslinger\ \cite{fermiexp}. A comparison of momentum distribution of the
superfluid fermions before and the after the dynamics could be used
to measure the momentum distribution of the defects generated during
the drive. Our theory predicts that this momentum distribution would
depend on the pairing symmetry of the superfluid and its pattern
would be qualitatively similar to that showing in \ \fref{fig6}
for $s-$ and $d_{x^2-y^2}-$wave superfluids.

\begin{figure}
\rotatebox{0}{\includegraphics*[width=\linewidth]{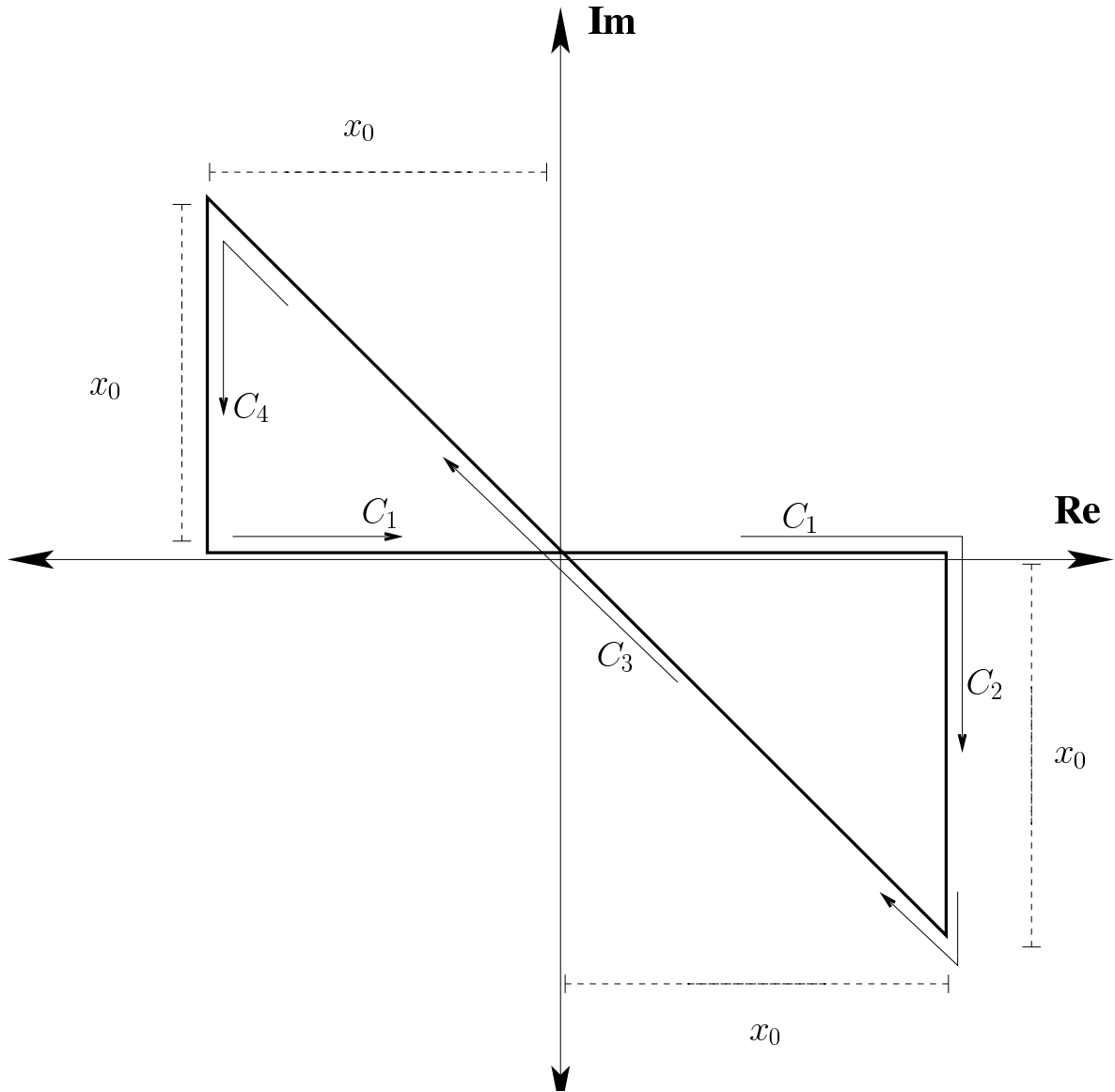}}
\caption{Contour
$\mathrm{C}=\mathrm{C}_1+\mathrm{C}_2+\mathrm{C}_3+\mathrm{C}_4$
used for evaluating the integral in   (\eref{eq:afk}) in the
complex $x$-plane. The path traversed is indicated by arrows, and
the contour consists of two isosceles right angled triangles of
height $x_0$ laid out as shown above.} \label{fig:lzcontour}
\end{figure}

In conclusion, we have studied the non-equilibrium periodic dynamics
of Fermi superfluids in the BCS regime within a self-consistent
mean-field theory. We have shown that proper incorporation of the
self-consistency condition is crucial for understanding the dynamic
properties of such systems. This is particularly highlighted by
studying the behaviour of the effective magnetization $m(t)$ (or
equivalently $n_0$) which shows qualitatively different behaviour for
Fermi superfluids (which obey self-consistent BCS equations) and
Ising or Kitaev spin models (whose properties are governed by
BCS-like equations without the self-consistency condition). We have
also studied the behaviour of defect density, it's momentum
distribution, and the residual energy for such dynamics. In
particular, we find that the momentum distribution of the defect
density bears a signature of the pairing symmetry of such
superfluids. Finally, we have provided an analytical derivation of
the frequency dependence $m_{F}$ at the end of one drive cycle in
the limit of large drive frequency and have shown that $m_{F} \sim
1/\omega$ for $\omega \gg \Delta_0$.

\ack{KS thanks DST, India for support under Project No.
SR/S2/CMP-$001/2009$. AR thanks CSIR, India for support under
Scientists' Pool Scheme No. $13(8531-$A$)/2011/$Pool.}

\appendix
\section{}
\label{appa}

Here we provide details of the evaluation of  \ \eref{eq:afk} in
\ \sref{analytic}. The integral in the right side of  \
\eref{eq:afk} can be evaluated by following the contour
$\mathrm{C}=\mathrm{C}_1+\mathrm{C}_2+\mathrm{C}_3+\mathrm{C}_4$ in
the complex plane as shown in \ \fref{fig:lzcontour}. Along the
path $\mathrm{C}_1$, we have $\mathrm{d}z=\mathrm{d}x$ and
$\exp{(-iz^2)}=\exp{(-ix^2)}$. Along the path $\mathrm{C}_2$, we
have $\mathrm{d}z=-i\mathrm{d}y$ and
$\exp{(-iz^2)}=\exp{[-i(x_0-iy)^2]}=\exp{[-i(x^2_0-y^2)]}\times
\exp{(-2x_0y)}$, an integrand that vanishes when
$x_0\rightarrow\infty$. Along the path $\mathrm{C}_3$, we have
$\mathrm{d}z=(1-i)\mathrm{d}x$ and $\exp{(-iz^2)}=\exp{(-2x^2)}$.
Finally, the integrand vanishes along path $\mathrm{ C}_4$ in a way
similar to that along $\mathrm{C}_2$. Since the contour $\mathrm{C}$
does not enclose any poles, Cauchy's theorem yields
$\oint_{\mathrm{C}}\mathrm{d}z\theta_{\mathbf k}(z)e^{-iz^2} =0$. Thus,
taking the limit $x_0\rightarrow\infty$,
\begin{eqnarray}
\label{eq:afkrhs}
\fl \int^\infty_{-\infty} \mathrm{d}x_{\mathbf k}\theta_{\mathbf k}(x_{\mathbf k}) e^{-ix^2_{\mathbf k}} = \nonumber\\
\left(1-i\right)\int^\infty_{-\infty} \mathrm{d}x_{\mathbf k}
\theta_{\mathbf k}\left[\left(1-i\right)x_{\mathbf k}\right] e^{-2x^2_{\bf
k}}.
\end{eqnarray}
Performing a Taylor expansion of $\theta_{\mathbf k}$,
\begin{equation}
\theta_{\mathbf k}(x_{\mathbf k}) = \sum^\infty_{n=0}\frac{x^n_{\bf
k}}{n!}\frac{\partial^n\theta_{\mathbf k}}{\partial^n x_{\bf
k}}\bigg|_{x_{\mathbf k}=0},
\end{equation}
and substituting this into the right side of  \ \eref{eq:afkrhs}
after the transformation $x_{\mathbf k}\rightarrow (1-i)x_{\mathbf k}$, each
term in the sum can be evaluated using Gaussian integrals, yielding
\begin{eqnarray}
\fl \int^\infty_{-\infty} \mathrm{d}x_{\mathbf k}\theta_{\mathbf k}\left[\left(1-i\right)x_{\mathbf k}\right] e^{-2x^2_{\mathbf k}} = \nonumber\\
\sqrt{\frac{\pi}{2}}\sum^\infty_{n=0} \frac{1}{n!}
\frac{\left(-i\right)^n}{\left(4\mu_a\omega\right)^n}
\frac{\partial^{2n}\theta_{\mathbf k}}{\partial^{2n}t}\bigg|_{t=t_{\bf
k}},
\end{eqnarray}
Using the above result, we evaluate  \ \eref{eq:afkrhs} and hence
 \ \eref{eq:afk}. This  yields   \eref{eq:afkfinal} used in\
 \sref{analytic}.

\end{document}